\documentclass[
  aps,
  pra,
  amsfonts,
  amsmath,
  amssymb,
  twocolumn,
  showpacs,
  floatfix,
  superscriptaddress
]{revtex4}
\usepackage{bm}
\usepackage{amsfonts}
\usepackage{amsmath}
\usepackage{epsfig}

\newcommand{\ave}[1]{\langle #1 \rangle}
\newcommand{\Figure}[1]{Figure~\ref{#1}}
\newcommand{\Fig}[1]{Fig.~\ref{#1}}
\newcommand{\Sec}[1]{Sec.~\ref{#1}}
\newcommand{\Figs}[1]{Figs.~\ref{#1}}
\newcommand{\Ref}[1]{Ref.~\cite{#1}}
\newcommand{\Refs}[1]{Refs.~\cite{#1}}
\newcommand{\Eq}[1]{Eq.~(\ref{#1})}

\newcommand{\Table}[1]{Table~\ref{#1}}
\newcommand{\mf}{{\text{mf}}}
\newcommand{\fit}{{\text{fit}}}
\newcommand{\med}{{\text{med}}}
\newcommand{\trap}{{\text{trap}}}
\newcommand{\ho}{{\text{ho}}}
\newcommand{\highT}{{\text{ht}}}
\newcommand{\tof}{{\text{tof}}}
\newcommand{\exc}{{\text{exc}}}
\newcommand{\osc}{{\text{osc}}}
\newcommand{\up}{\uparrow}
\newcommand{\down}{\downarrow}
\newcommand{\spin}{s}
\newcommand{\eq}{\text{eq}}
\newcommand{\fs}{f_\spin}
\newcommand{\fsp}{f_\spin^{\prime}}
\newcommand{\fup}{f_{\up}}
\newcommand{\fupp}{f_{\up}^{\prime}}
\newcommand{\fdo}{f_{\down}}
\newcommand{\fdop}{f_{\down}^{\prime}}
\newcommand{\fz}{f_0}

\newcommand{\mubar}{\bar{\mu}}

\newcommand{\wbar}{\bar{w}}
\newcommand{\omr}{\omega_r}
\newcommand{\omx}{\omega_x}
\newcommand{\omy}{\omega_y}
\newcommand{\omz}{\omega_z}
\newcommand{\ombar}{\bar{\omega}}
\newcommand{\vek}[1]{\bm{\mathrm{#1}}}
\newcommand{\nablav}{\bm{\nabla}}
\newcommand{\rv}{\vek{r}}
\newcommand{\pv}{\vek{p}}
\newcommand{\ps}{\vek{p}_\spin}
\newcommand{\pupv}{\vek{p}_{\up}}
\newcommand{\pdov}{\vek{p}_{\down}}

\newcommand{\kv}{\vek{k}}
\newcommand{\qv}{\vek{q}}
\newcommand{\Fv}{\vek{F}}
\newcommand{\vv}{\vek{v}}

\newcommand{\kpqp}{\kv/2+\qv'}

\newcommand{\kmqp}{\kv/2-\qv'}

\newcommand{\dthreep}{\frac{d^3p}{(2\pi)^3}}
\newcommand{\dthreepup}{\frac{d^3p_{\up}}{(2\pi)^3}}
\newcommand{\dthreepdo}{\frac{d^3p_{\down}}{(2\pi)^3}}

\newcommand{\dthreeqp}{\frac{d^3q'}{(2\pi)^3}}
\newcommand{\omQ}{\omega_{\mathrm{Q}}}
\newcommand{\GamQ}{\Gamma_{\mathrm{Q}}}

\DeclareMathOperator{\Real}{Re}

\DeclareMathOperator{\atanh}{atanh}
\DeclareMathOperator{\maximum}{max}
\begin{document}
\title{Numerical solution of the Boltzmann equation for trapped Fermi
  gases with in-medium effects}
\author{Pierre-Alexandre Pantel}
\affiliation{Universit{\'e} de Lyon, Univ. Lyon 1, CNRS/IN2P3,
IPN Lyon, F-69622 Villeurbanne Cedex, France}
\author{Dany Davesne}
\affiliation{Universit{\'e} de Lyon, Univ. Lyon 1, CNRS/IN2P3,
IPN Lyon, F-69622 Villeurbanne Cedex, France}
\author{Michael Urban}
\affiliation{Institut de Physique Nucl{\'e}aire, CNRS-IN2P3 and
Universit\'e Paris-Sud, 91406 Orsay Cedex, France}
%
\begin{abstract}
Using the test-particle method, we solve numerically the Boltzmann
equation for an ultra-cold gas of trapped fermions with realistic
particle number and trap geometry in the normal phase. We include a
mean-field potential and in-medium modifications of the cross-section
obtained within a T matrix formalism. After some tests showing the
reliability of our procedure, we apply the method to realistic cases
of practical interest, namely the anisotropic expansion of the cloud
and the radial quadrupole mode oscillation. Our results are in good
agreement with experimental data. Although the in-medium effects
significantly increase the collision rate, we find that they have only
a moderate effect on the anisotropic expansion and on frequency and
damping rate of the quadrupole mode.
\end{abstract}
%
\pacs{67.85.Lm,02.70.Ns}
\maketitle

\section{Introduction}
Ultracold gases of trapped atoms offer unique opportunities to study
Fermionic many-body systems under clean and adjustable conditions. In
particular, by tuning the magnetic field around a Feshbach resonance,
one can realize either a repulsive or an attractive interaction,
passing through the ``unitary limit'' where the scattering length $a$
diverges. Especially this ``unitary Fermi gas'' has attracted a lot of
attention because it has ``universal'' properties (there is no scale
depending on the interaction strength) and it has some similarities
with completely different systems such as low-density neutron
matter. A very interesting aspect of these strongly correlated atomic
gases are the different dynamical regimes they can be in: they can be
superfluid or normal-fluid, and in the normal-fluid phase hydrodynamic
to collisionless behavior can be realized depending on interaction
strength and temperature. Experimentally, the transition between these
regimes has been seen, e.g., by studying the frequency and damping
rates of the radial quadrupole mode
\cite{Wright2007,Riedl2008}. Another source of information is the free
expansion (i.e., after the trap is switched off) of an anisotropic
gas, which is very different in the ballistic and in the hydrodynamic
regime
\cite{OHara2002,Menotti2002,Pedri2003,Schaefer2010}. Surprisingly, the
low viscosity of the unitary Fermi gas observed in this way
\cite{Cao2011,Elliott2014a,Elliott2014b,Joseph2014} seems to indicate
some analogies even with the quark-gluon plasma created in
ultrarelativistic heavy-ion collisions \cite{Schaefer2009}.

In the normal phase, the transition between hydrodynamic and
collisionless regimes can be described by the Boltzmann equation
\cite{Toschi2003,Massignan2005,BruunSmith2007_scissors,Riedl2008,Chiacchiera2009,Lepers2010,Chiacchiera2011,WuZhang2012}. The
test-particle method for the numerical solution of the Boltzmann
equation has been used for many years in the field of heavy-ion
collisions \cite{Bertsch1988} and more recently also for ultracold
atoms
\cite{Toschi2003,Toschi2004,Lepers2010,Wade2011,Goulko2011,Goulko2012}. In
this article, we extend the method developed in \cite{Lepers2010} in
order to simulate realistic systems (trap geometry and particle
number). At the same time, we include in-medium effects into the
calculation, namely the ``mean-field'' potential
\cite{Chiacchiera2009,Pantel2012} and the in-medium cross-section
\cite{Riedl2008,BruunSmith2005,Chiacchiera2009,BluhmSchaefer2014},
obtained within a T matrix theory.

In \Sec{sec:methods}, we briefly recall the Boltzmann equation with
in-medium effects and explain the test-particle method we use to solve
it numerically, concentrating mainly on elements which are new as
compared to \Ref{Lepers2010}. Then we describe in \Sec{sec:tests} some
tests of the reliability of the code. In \Sec{sec:results}, we discuss
as first applications of the code the simulation of different
anisotropic expansion experiments of the Duke group
\cite{Cao2011,Elliott2014a} and the radial quadrupole mode as measured
at Innsbruck \cite{Riedl2008}. In \Sec{sec:conclusion} we conclude and
discuss perspectives for future work.

Throughout the article, we use units with $\hbar = k_B = 1$ ($\hbar =$
reduced Planck constant, $k_B = $ Boltzmann constant).
\section{Methods}\label{sec:methods}
\subsection{Boltzmann equation}
We study a two-component $(\spin \in\{\up,\down\}$) ultra-cold gas of
$N = N_\up+N_\down$ atoms of mass $m$ with attractive interaction
(scattering length $a<0$) in an external potential $V_\trap(\rv,t)$.
Even if we will restrict the numerical implementation of the in-medium
effects to the case of a spin balanced ($N_\up = N_\down$) system, we
keep the notation general. We will only consider the normal-fluid
phase, i.e., temperatures $T$ above the superfluid transition
temperature $T_c$. In this case, the dynamics of the system can be
described with a distribution function $\fs(\rv,\pv,t)$ which
satisfies the Boltzmann equation
\begin{equation}
  \label{eq:boltzmann}
  \frac{\partial \fs}{\partial t} + \frac{\pv}{m} \cdot \frac{\partial
  \fs}{\partial \rv} + \Fv_\spin(\rv,t) \cdot \frac{\partial \fs}{\partial
  \pv} = -I_\spin\,,
\end{equation}
where $I_\spin$ is the collision integral (see below) and
$\Fv_\spin = \Fv_\trap + \Fv_{\mf,\spin}$ is the sum of the forces
resulting from the external potential $V_\trap(\rv,t)$ and from the
mean field $U_\spin(\rv,t)$.

The trap potential $V_\trap$ is essentially given by a time
independent anisotropic harmonic trap
\begin{equation}
  V_\trap(\rv) = \frac{1}{2} m (\omx^2 x^2 + \omy^2 y^2 + \omz^2 z^2 )\,,
\end{equation}
(in the cylindrically symmetric case, we use $\omr = \omx = \omy$)
which defines the time and length scales $\ombar^{-1} = (\omx \omy
\omz)^{-1/3}$ and $\bar{L}_\ho = (m\ombar)^{-1/2}$. In addition
to the static harmonic potential, $V_\trap$ can have a time dependent
contribution, e.g., in order to excite a collective mode ($V_\exc$).

The density per spin state and the number of atoms of species $\spin$
are obtained from the distribution function as
\begin{align}
  \rho_\spin(\rv,t) &= \int \dthreep \fs(\rv,\pv,t)\,, &
  N_\spin &= \int d^3r \rho_\spin(\rv,t)\,.
\end{align}

The right-hand side (r.h.s.) of the Boltzmann equation
(\ref{eq:boltzmann}) describes the collisions between particles of
opposite spin. It thus depends on the differential scattering cross
section $d\sigma/d\Omega$. If we consider, e.g., the $\up$ component,
it reads explicitly:
\begin{multline}
  I_\up 
    = \int \dthreepdo \int d\Omega \frac{d\sigma}{d\Omega}
      \frac{|\pupv - \pdov|}{m} \\
      \times \left[ \fup \fdo (1-\fupp) (1 - \fdop) 
        - \fupp \fdop (1 - \fup) (1 - \fdo) \right],
\end{multline}
where, in the first term, $\ps$ and $\ps'$ are the incoming and
outgoing momenta, respectively, while in the second term, the roles
are exchanged. We adopted the short-hand notation $\fs =
\fs(\rv,\ps,t)$ and $\fsp = \fs(\rv,\ps',t)$.

Since we consider only $s$-wave interactions, the cross section does
not depend on the scattering angle $\Omega$ and we simply have
$\sigma = 4\pi\, d\sigma/d\Omega$. In free space, it depends only on
the relative momentum $q= |\pupv - \pdov|/2$ and is given by
\begin{equation}
  \sigma = \sigma_0(q) = \frac{4 \pi a^2}{1 + (qa)^2}.
\end{equation}
When in-medium effects are taken into account (see
\Sec{ss:inmed_xsec}), there is of course no such simple form.

\subsection{Test-particle method}\label{subsec:testparticles}
In this article, we use the test-particle method described, e.g., in
\cite{Lepers2010}, but generalized to the case $\nu < 1$, where
$\nu = \tilde{N}/N$ is the number of test particles per atom. This
case is of course of practical interest since it allows us to simulate
a large number $N$ of real particles with a smaller number $\tilde{N}$ of
test particles. However, the problem of a correct sampling of the
phase space immediately arises.

As discussed in detail in \cite{Lepers2010}, the main problem when
reducing the number of test particles is that one cannot resolve any
more the step of the distribution function at the Fermi surface. This
is, however, crucial for the description of Pauli blocking of
collisions. The resolution can be reduced at finite temperature since
the Fermi surface gets smoothed out. In the unitary limit, the
superfluid transition temperature $T_c$ is already of the order of
$\sim 0.3\,T_F$, where $T_F$ is the Fermi temperature defined in the
usual way by $T_F = (3N)^{1/3}\ombar$ (which is lower than the local
Fermi temperature in the center of the trap). Since we are anyway
forced to stay above $T_c$, we can work with a smaller number of test
particles.

The test-particle method consists of writing the (non-equilibrium)
distribution function as
\begin{equation}
  \fs(\rv,\pv,t) = \frac{1}{\nu} \sum_{i_\spin} \delta(\rv -
  \rv_{i_\spin}(t)) \delta(\pv - \pv_{i_\spin}(t)),
\label{ftestparticle}
\end{equation}
where $\rv_i$ and $\pv_i$ are the position and momentum of the $i$-th
test particle at time $t$. In practice, the first $\tilde{N}_\up =
\nu N_\up$ test particles represent atoms with spin $\up$, while
the remaining $\tilde{N}_\down = \nu N_\down$ test particles
represent atoms with spin $\down$. To simplify the notation, we use
the convention that the index $i_\up$ runs from $1$ to $\tilde{N}_\up$,
the index $i_\down$ from $\tilde{N}_\up+1$ to $\tilde{N}$, and the index $i$
from $1$ to $\tilde{N} = \tilde{N}_\up+\tilde{N}_\down$.

With the test-particle method, the l.h.s. of the Boltzmann equation is
replaced by a set of $6 \tilde{N}$ classical equations of motion
\begin{equation} 
  \dot{\rv}_{i_\spin}(t) = \frac{\pv_{i_\spin}(t)}{m}\,, \quad
  \dot{\pv}_{i_\spin}(t) = \Fv_\spin(\rv_{i_\spin}(t),t)\,.
\end{equation}
To solve these equations, a particular attention has to be paid to the
efficiency and numerical accuracy. For our simulations, we use the
velocity Verlet algorithm as in \cite{Lepers2010}.

The collision term is simulated by scattering two test particles of
opposite spin by a random angle if their distance becomes smaller than
that corresponding to the cross section (see \cite{Lepers2010} for
details). To take Pauli blocking into account, the collision is only
performed with a probability given by $(1-\fupp)(1-\fdop)$.

Of course, the $\delta$ functions in \Eq{ftestparticle} cannot be used
to obtain meaningful numbers for the Pauli-blocking factors
$(1-\fupp)(1-\fdop)$. The usual way to avoid this problem is to
consider Gaussian distributions in $\rv$ and $\pv$ spaces instead of
the Dirac ones. In the calculation of the Pauli-blocking factors, we
therefore replace $\delta(\rv-\rv_i) \delta(\pv-\pv_i)$ with $
g_r(\rv-\rv_i) g_p(\pv-\pv_i)$, where
\begin{equation}
  g_p(\pv) = \frac{\exp(-p^2/w_p^2)}{(\sqrt{\pi}w_p)^3},
\end{equation}
and
\begin{equation}
  g_r(\rv) = \frac{\exp(-x^2/w_{r,x}^2 - y^2/w_{r,y}^2 - z^2/w_{r,z}^2)}
    {(\sqrt{\pi}\wbar_r)^3} .
\end{equation}
In contrast to the Gaussian used in \cite{Lepers2010}, the above
Gaussian $g_r$ takes explicitly into account the deformed shape of the
gas. To be specific, the widths appearing in the above equation are
defined from an average width $\wbar_r$ as
\begin{equation}
  w_{r,j} = \frac{\ombar}{\omega_j} \wbar_r,
\label{differentwidths}
\end{equation}
where $j \in \{x,y,z\}$. In this way we exploit the fact that the
typical length scales on which the distribution function varies, given
by the trap potential, can be very different in the three spatial
directions.

The values of $\wbar_r$ and $w_p$ must be chosen such that they smooth
the statistical fluctuations due to the finite number of
test particles but resolve the structure of the distribution function
$\fs$ (see \cite{Lepers2010} for details). The parameters we use in
the present paper are given in \Table{tab:numericalparameters}.
\begin{table}
\caption{Numerical parameters used in the simulations shown in the
  figures.
  \label{tab:numericalparameters}}
   \begin{ruledtabular}
   \begin{tabular}{cccc}
      Figs.&&\ref{fig:kohn_mf}, \ref{fig:test_cr}, 
        \ref{fig:quad}--\ref{fig:quad_exp} & 
        \ref{Fig:expansion}--\ref{Fig:compareCao}\\
      $\tilde{N}$ &                   & $5\cdot 10^4$ & $1\cdot 10^5$\\
      $\Delta t$  & ($\ombar^{-1}$)    & $0.02$        & $0.01$\\
      $\wbar_r$   & ($\bar{L}_\ho$)    & $3$           & $3$\\
      $w_p$       & ($\bar{L}_\ho^{-1}$)& $3$           & $3$\\
      $\wbar$     & ($\bar{L}_\ho$)    & $2$           & $2$
   \end{tabular}
   \end{ruledtabular}
\end{table}

Another improvement w.r.t. \Ref{Lepers2010} concerns the inclusion of
the in-medium cross section $\sigma$ and of the mean field
$U_\spin$. For simplicity, we assume that these quantities can be
calculated at a local equilibrium characterized by time-dependent
effective local chemical potentials and temperature,
$\mu^*_\spin(\rv,t)$ and $T^*(\rv,t)$. In practice, it is more
convenient to parameterize $\sigma$ and $U$ in terms of the densities
$\rho_\spin$ and of the kinetic-energy density, $\epsilon$, which are
related by a one-to-one correspondence to $\mu^*_\spin$ and $T^*$:
\begin{gather}
\rho_\spin = \int \frac{d^3k}{(2\pi)^3}
  \frac{1}{e^{(k^2/2m+U_\spin-\mu^*_\spin)/T^*}+1}\,,\\ 
\epsilon = \sum_\spin \int \frac{d^3k}{(2\pi)^3}
  \frac{k^2/2m}{e^{(k^2/2m+U_\spin-\mu^*_\spin)/T^*}+1}\,.
\end{gather}
When one calculates $\rho_\spin$ and $\epsilon$ from the
test-particle distribution, it is enough to introduce an averaging in
$\rv$ space (contrary to the case of the Pauli-blocking factors
discussed above, where also averaging in $\pv$ space is
needed). Again, we use Gaussians for this purpose, with widths
$w_j$ which are derived from an average width $\bar{w}$ analogously to
\Eq{differentwidths}. The width $\bar{w}$ can be chosen smaller than
the width $\bar{w}_r$ used in the Pauli-blocking factors, because in
the density and energy density we sum over all momenta and therefore
the statistical fluctuations are smaller. During the simulation, the
averaged densities and kinetic-energy density are computed as
\begin{gather}
\tilde{\rho}_\spin(\rv,t) = \frac{1}{\nu} \sum_{i_\spin}
  g(\rv-\rv_{i_\spin})\,,\\
\tilde{\epsilon}(\rv,t) = \frac{1}{\nu} \sum_i
  g(\rv-\rv_i) \frac{[\pv_i-m\tilde{\vv}(\rv,t)]^2}{2m}\,. 
\end{gather}
Note that $\epsilon$ is defined in the frame moving with the
local average velocity
\begin{equation}
  \tilde{\vv}(\rv,t) = \frac{1}{\nu \tilde{\rho}(\rv,t)}
  \sum_i g(\rv-\rv_i) \frac{\pv_i}{m}\,,
\end{equation}
where $\tilde{\rho} = \tilde{\rho}_\up + \tilde{\rho}_\down$.
These quantities are calculated after every time step and stored on a
three-dimensional (3d) grid.

Since the mean fields $U_\spin$ are now functions of the smoothed
densities $\tilde{\rho}_\spin$, one would violate Newton's third law
if one computed the corresponding force directly as
$\Fv_{\mf,\spin} = -\nablav U_\spin$. Following
\cite{Urban2012}, one has to average the force using the same
Gaussians $g$ as in the calculation of $\tilde{\rho}_\spin$,
i.e., the contribution to the force on a test particle $i_\spin$
caused by its interactions with all the other test particles is given
by
\begin{multline}
\Fv_{\mf,\spin}(\rv_{i_\spin},t) = -\frac{\partial}{\partial \rv_{i_\spin}}
  \tilde{U}_\spin(\rv_{i_\spin},t)\\
  = -\frac{\partial}{\partial \rv_{i_\spin}} \int d^3r 
  g(\rv_{i_\spin}-\rv) U_\spin(\rv,t)\,.
\end{multline}
It is straight-forward to show that with this prescription the sum of
all interaction forces would be exactly zero, as required by momentum
conservation, if the mean fields $U_\spin$ were only functions of the
densities $\tilde{\rho}_\spin$. However, since the mean fields depend
also on $\tilde{\epsilon}$, this relation is probably only
approximately fulfilled (see also \Sec{sec:Kohn}).

\subsection{In-medium cross section}
\label{ss:inmed_xsec}
Since the cross section governs the collisions, the introduction of
in-medium effects may have important consequences in the dynamics of
the system. Following \cite{Chiacchiera2009}, where a T-matrix
approximation is used to calculate the in-medium correlations, we
write the expression for the cross section as
\begin{equation}
  \label{eq:inmedxsec}
  \frac{d\sigma_\med(\kv,\qv)}{d\Omega} = \left| \frac{m}{4\pi}\Gamma(\omega,\kv)
  \right|^2 ,
\end{equation}
where $\omega = k^2/4m + q^2/m-2\mubar$ is the total energy
(w.r.t. $2\mubar = \mu_\up+\mu_\down$), $\kv = \pv_\up+\pv_\down$ and
$\qv = (\pv_\up-\pv_\down)/2$ are total and relative momenta of the
colliding particles with spin $\up$ and $\down$. The in-medium T
matrix $\Gamma$ is defined as
\begin{equation}
  \Gamma(\omega,\kv) = \frac{4\pi a}{m}\,
    \frac{1}{1 + i a q - \frac{4\pi a}{m}J(\omega,\kv)}\,,
\end{equation}
with $q = \sqrt{m(\omega+2\mubar)-k^2/4}$. The medium
contribution to the non-interacting two-particle Green's function
reads
\begin{equation}
  J(\omega,\kv) = -\int \dthreeqp \frac{\fz(\xi_{\kpqp}^\up) +
  \fz(\xi_{\kmqp}^\down)}{\omega - \xi_{\kpqp}^\up - \xi_{\kmqp}^\down +
  i \eta}.
\label{eq:J}
\end{equation}
In the above equation, $\fz(\xi) = 1/(e^{\beta\xi}+1)$ is the Fermi
function with $\beta = 1/T$, and $\xi_{\kv}^\spin = k^2/2m - \mu_\spin$
represents the single-particle energy (measured w.r.t. the chemical
potential $\mu_\spin$) of an atom with spin $\spin$. As usual, the
limit $\eta \to 0$ is implicit.

This approach has already been used in \Refs{Chiacchiera2009,
  Chiacchiera2011} to study collective modes in an unpolarized gas
($N_\up = N_\down$). Actually, since the standard (non
self-consistent) T-matrix approximation fails in the polarized case,
we will include the medium effects only in the unpolarized case.

Strictly speaking, the in-medium cross section given above is only
valid for a gas described by a thermal equilibrium distribution $\fz$,
characterized by chemical potentials $\mu_\spin$ and a temperature
$T$. As mentioned in the preceding subsection, we use this cross
section also in the non-equilibrium case, calculating it with the
local effective chemical potentials $\mu^*_\spin$ and effective
temperature $T^*$ which give the same densities $\rho_\spin$ and
energy density $\epsilon$ as the actual (non-equilibrium) distribution
function $f$. While this is exact in the linear response regime, it is
of course only an approximation in situations far from equilibrium.

Including in-medium effects in a simulation implies a huge increase of
the computation time: in order to decide whether a collision between
an $\up$ and a $\down$ particle is allowed, the first criterion is
given by the relative distance (and thus the cross-section
$\sigma_\med$) between the test particles. Therefore, $\sigma_\med$
(which depends on the relative and total momenta of the two colliding
test particles) has to be determined $\tilde{N}_\up \tilde{N}_\down$
times at each time step. The computation of the real part of $J$
involves a numerical integration \cite{Chiacchiera2009}, and therefore
it would be much too time-consuming to include the exact in-medium
cross section into the simulation. It is clearly necessary to use a
simple parameterization of the cross section which can be evaluated
quickly. More details are given in Appendix \ref{app:xsection}.
\subsection{Mean-field potential}
\label{ss:mean_field}
In order to be consistent, we also have to take into account the
in-medium effects in the left-hand side of the Boltzmann equation,
i.e., the mean-field potential $U$. Following \Ref{Chiacchiera2009},
we calculate it as
\begin{equation}
  U = \Real \Sigma(0,k_{\mubar^*})\,,
\end{equation}
where $\Sigma$ is the self-energy in ladder approximation
$k_{\mubar^*} = \sqrt{2m\maximum(\mubar^*,0)}$ (see
\Ref{Chiacchiera2009} for further details). Like the cross section,
this mean-field depends on $\rv$ and $t$ through the effective
chemical potential $\mubar^*$ and temperature $T^*$ determined from
the density $\bar{\rho}$ [since we can calculate the mean field only
  in the unpolarized case, we use $\bar{\rho} =
  (\rho_\up+\rho_\down)/2$] and energy density $\epsilon$. In
practice, for a given $a$, we tabulate it directly as function of
$\bar{\rho}$ and $\epsilon$, using the Nozi\`eres-Schmitt-Rink relation between
$\bar{\rho}$ and $\mubar^*$ (cf. Ref. [15]).

\subsection{Initialization}
The last point concerns the initialization process. The system is
initialized according to an equilibrium Fermi distribution (the same
for $\up$ and $\down$ particles since we limit ourselves to balanced
systems for the moment) for a given temperature $T$
\begin{equation}
  f_{\eq,\spin}(\rv,\pv) = \frac{1}{e^{\left[ p^2/2m + V_\trap(\rv) +
        \tilde{U}(\rv) - \mu\right]/T} + 1}.
\label{eq:finitial}
\end{equation}
Note that the mean field $\tilde{U}$ entering here is the one folded
with the Gaussian $g$ as described in
\Sec{subsec:testparticles}, and it has to be calculated
self-consistently from the folded density and energy density
$\tilde{\rho}$ and $\tilde{\epsilon}$, otherwise the initial
distribution would not be stationary under the time evolution of the
code \cite{Urban2012}.

In practice, we store $\rho$, $\tilde{\rho}$, $\epsilon$,
$\tilde{\epsilon}$, $U$, and $\tilde{U}$ on a 3d
grid. Starting from the density profile of an ideal Fermi gas, we
iterate them simultaneously with $\mu$ in order to converge to a
self-consistent solution with the correct particle number. In the end,
because of the attractive mean field, we obtain a density profile that
is more concentrated in the trap center (cf.\ Fig.~3 of
\Ref{Chiacchiera2009}) and a chemical potential that is lower than
that of the ideal Fermi gas.

Once $\mu$ and $\tilde{U}$ are determined, the positions $\rv_i$ and
momenta $\pv_i$ are initialized randomly according to the distribution
(\ref{eq:finitial}), as in \Ref{Lepers2010}.
\section{Tests of reliability and accuracy}
\label{sec:tests}
\subsection{Sloshing mode}
\label{sec:Kohn}
The first test to check the consistency of our numerical approach
concerns the particle propagation (left-hand side of the Boltzmann
equation). This has already been done in \Ref{Lepers2010} but without
mean-field. Therefore, we will concentrate on the implementation of
$U(\rv,t)$. One conclusive way to check this is to excite the sloshing
mode (center-of-mass oscillation). As mentioned in \cite{Pantel2012},
the mean-field potential must not have any effect on the sloshing
frequency when the trapping potential is harmonic. For instance, if
the mode along the $x$ direction is excited, the frequency of the
sloshing mode must be $\omx$ (Kohn's theorem \cite{Kohn1961}).

\begin{figure}
  \includegraphics[width=8cm]{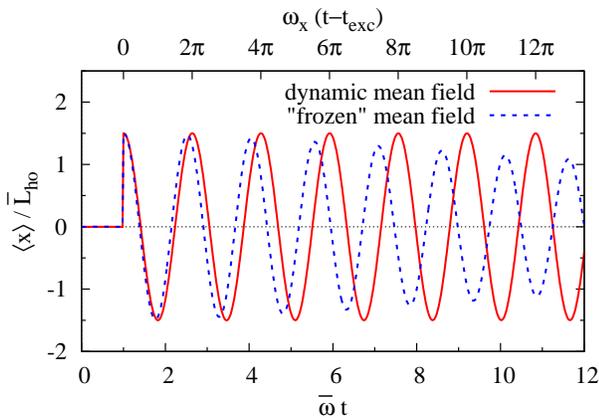}
  \caption{Simulation of the sloshing mode in $x$ direction in a
    realistic system whose parameters are given in
    \Table{tab:expinnsbruck} at $1/k_Fa = -0.1$ and $T/T_F =
    0.4$. The result of the full calculation (solid line) is compared
    with the result obtained with a ``frozen'' mean field (dashed
    line, see text).
  \label{fig:kohn_mf}}
\end{figure}

Figure~\ref{fig:kohn_mf}
shows the result for the sloshing mode for a realistic system whose
parameters are listed in \Table{tab:expinnsbruck}.
\begin{table}
\caption{Parameters of the quadrupole mode experiment of
  \Ref{Riedl2008}.\label{tab:expinnsbruck}}
   \begin{ruledtabular}
   \begin{tabular}{ccc}
      $\omr/2\pi$& (Hz) & 1800\\
      $\omz/2\pi$& (Hz) & 32\\
      $N$  & & $6\cdot 10^5$
   \end{tabular}
   \end{ruledtabular}
\end{table}
The excitation has been produced at $\ombar t_\exc = 1$ by a global
shift of all particle positions by $1.5 \bar{L}_\ho$ in $x$ direction.
Note that this shift is comparable to the cloud width $\sqrt{\ave{x^2}}
\approx 1.8 \bar{L}_\ho$, i.e., we are no longer in the linear response
regime. The solid line corresponds to the full simulation, which takes
the mean field and the in-medium cross section into account. We
observe an undamped oscillation of the center of mass $\ave{x}$ with
frequency $\omega_x$ (cf. upper scale in \Fig{fig:kohn_mf}), in
perfect agreement with Kohn's theorem.

In order to show that this is indeed a non-trivial test of the
calculation of the mean field $U(\rv,t)$, we show for comparison as
the dashed line the result one obtains if the mean field is ``frozen''
(i.e., not changed during the time evolution) at its initial value
$U(\rv,t=0)$. In this case, the oscillation of the cloud is faster
than $\omega_x$ and damped, because the total potential $V_\trap+U$ is
deeper than $V_\trap$ and anharmonic.

We thus have demonstrated that the particle propagation in the
presence of the mean field, and the calculation of the mean field from
the distribution function, are well described in our code.
\subsection{Collision rate}
A good way to check the precision of the simulation of the collision
term is to compare the collision rate per particle in equilibrium
given by the code with its exact value obtained directly from the
definition
\begin{multline}
  \gamma = \frac{1}{N} \int d^3r \int \dthreepup
  \int \dthreepdo \int d\Omega \frac{d\sigma}{d\Omega} \frac{|\pupv -
    \pdov|}{m}\\ \times \fup \fdo (1-\fupp) (1 - \fdop)
\label{eq:collrate}
\end{multline}
by Monte-Carlo integration. Actually, two ingredients are tested
simultaneously: firstly, the collision process including Pauli
blocking itself, and secondly, the parameterization of the in-medium
cross section.

\begin{figure}
  \includegraphics[width=8cm]{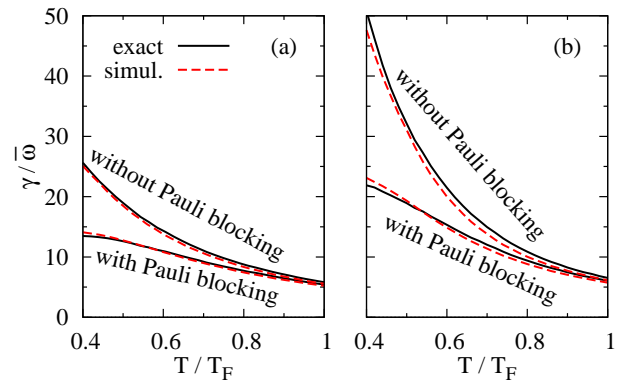}
  \caption{Temperature dependence of the collision rates with (lower
    curves) and without (upper curves) Pauli blocking for the same
    system as in \Fig{fig:kohn_mf}. (a) without in-medium effects,
    (i.e., $\sigma = \sigma_0$ and $U=0$); (b) with in-medium
    cross-section $\sigma_\med$ and mean field $U$.
  \label{fig:test_cr}}
\end{figure}

The lower curves in \Fig{fig:test_cr}
show the collision rates $\gamma$ obtained with the Boltzmann code
(dashed lines) and from \Eq{eq:collrate} (solid lines) as a function
of temperature. In order to check the cross section independently from
Pauli blocking, we also show the collision rates without Pauli
blocking, i.e., the sum of the rates of allowed and blocked collisions
[in this case, the exact result is obtained by integrating
  \Eq{eq:collrate} without the Pauli-blocking factors
  $(1-f'_\spin)$]. Two cases are examined: (a) without medium
effects (i.e., with the free cross section $\sigma_0$ and without mean
field) and (b) with in-medium effects (i.e., with the in-medium cross
section $\sigma_\med$ and with mean-field $U$). First, we can see on
panel (a) that without in-medium effects the collision rates are well
reproduced even at low temperature where the Pauli blocking is very
important. Looking now at panel (b), we observe that the agreement is
still satisfactory in the whole temperature range. Note that
the Monte-Carlo integrations have been performed with the exact
$\sigma_\med$, while in the simulation of course the approximated one
(cf. Appendix \ref{app:xsection}) was used. Thus, we can consider that
the approximation of $\sigma_\med$ is quite accurate and the collision
process is well described in the simulation. Finally, the comparison
between panels (a) and (b) shows that the in-medium effects
significantly increase the collision rate at low temperature, as
expected from previous calculations \cite{BruunSmith2005,Riedl2008}.

\section{Comparison with experiments}\label{sec:results}
The aim is now to apply our numerical solution of the Boltzmann
equation to realistic cases. We will compare our results with
different expansion and collective-mode experiments by the Duke and
Innsbruck groups.
\subsection{Anisotropic expansion}
The expansion of the gas from an anisotropic trap was shown to be a
useful tool to distinguish the collisionless (weakly interacting) case
from the hydrodynamic (strongly interacting) one
\cite{OHara2002,Menotti2002,Pedri2003}. Recently, more refined
analyses of the anisotropic expansion based on viscous hydrodynamics
\cite{Schaefer2010} were used to extract the temperature dependence of
the shear viscosity $\eta$ of the unitary Fermi gas from the
experimental expansion data
\cite{Cao2011,Elliott2014a,Elliott2014b,Joseph2014}. However, a
problem of these analyses is that near the surface of the cloud, the
gas is always in the collisionless regime, in which the concept of
viscous hydrodynamics is not applicable. It is therefore important to
analyse these experiments within the Boltzmann framework, which is
capable of describing the hydrodynamic expansion in the center of the
cloud and the ballistic expansion of the dilute outer region.

Here, we simulate four expansion experiments of Fermi gases in the
unitary limit whose parameters are listed in \Table{tab:expduke}.
\begin{table}
\caption{Parameters of the anisotropic expansion experiments of
  \Refs{Cao2011,Elliott2014a}.\label{tab:expduke}}
   \begin{ruledtabular}
   \begin{tabular}{cccccc}
      Ref. & & \cite{Elliott2014a} & \cite{Cao2011} &
        \cite{Cao2011} & \cite{Cao2011}\\
      $\omx/2\pi$& (Hz) & 2210 & 5283 & 5283 & 5283\\ 
      $\omy/2\pi$& (Hz) & 830  & 2052 & 5052 & 5052\\
      $\omz/2\pi$& (Hz) & 64.3 & 182.7 & 182.7 & 182.7\\
      $\omega_{\text{mag}}/2\pi$& (Hz) & 21.5 & 21.5 & 21.5 & 21.5\\
      $N$  & & $2.5\cdot 10^5$ & $4\cdot 10^5$ & $5\cdot 10^5$ & $6\cdot 10^5$\\
      $\tilde{E}/E_F$ & & $ 1.46 $ & $2.3$ & $3.3$ & $4.6$\\
      $T/T_F$ & & $0.51$ & $0.79$ & $1.11$ & $1.54$
   \end{tabular}
   \end{ruledtabular}
\end{table}
The quantity $\tilde{E}$ used by the Duke group to characterize the
temperature of the cloud (before the expansion) is twice the potential
energy per particle. We transform $\tilde{E}$ into temperature $T$ by
calculating $\tilde{E}$ with our equilibrium density profile including the
mean field $U$. The system is initialized with the given parameters in
the anisotropic trap, then the trap potential is switched off except
for the weak magnetic potential $V_\text{mag} = \tfrac{1}{2}
m\omega_{\text{mag}}^2 (y^2+z^2-2x^2)$ which was present in the
experiments during the expansion and which affects merely the
expansion in $z$ direction.

\begin{figure*}
  \includegraphics[width=12cm]{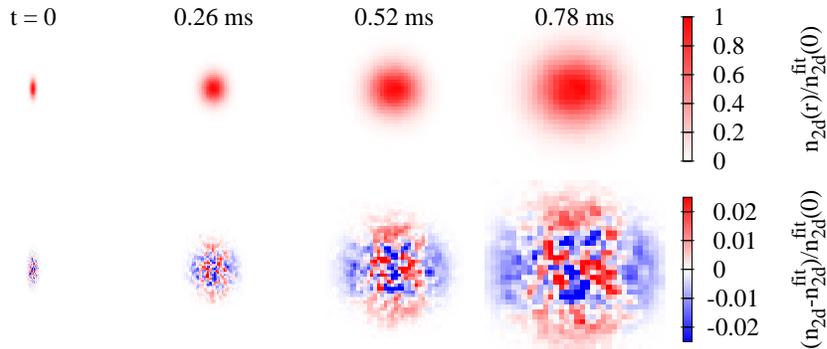}
  \caption{Simulation of the expansion experiment by Elliott et
    al. \cite{Elliott2014a} with $\tilde{E}/E_F = 1.46$. The upper row
    shows the column density $n_{2d}$ (i.e., the density integrated
    over $z$) in the $xy$ plane ($x$ in horizontal and $y$ in vertical
    direction) for different times $t$ after the beginning of the
    expansion. The lower row shows the error of the fit of the density
    profile with a Gaussian (see text).
    \label{Fig:expansion}}
\end{figure*}

In the upper row of \Fig{Fig:expansion}
we display the column density $n_{2d}(x,y) = \int dz\,n(x,y,z)$ in the
$xy$ plane at different times $t$ after the beginning of the
expansion, for initial conditions corresponding to the experiment of
\Ref{Elliott2014a} with $\tilde{E}/E_F = 1.46$. One can clearly see
the inversion of the aspect ratio during the expansion. To quantify
this, we apply the same procedure as in the analysis of the experiment
\cite{Elliott2014a} and determine the cloud widths $\sigma_x$ and
$\sigma_y$ by fitting the density profile with a Gaussian
\begin{equation}
n_{2d}^{\text{fit}}(x,y) \propto e^{-x^2/\sigma_x^2-y^2/\sigma_y^2}\,.
\end{equation}
The error of this fit is shown in the lower row of
\Fig{Fig:expansion}, it does not exceed $2.5\%$ of the central
density. In the center, one sees the numerical noise coming from the
finite number of test particles ($10^5$) used in the
simulation. However, in the peripheral regions, where the density is
low, we see systematic deviations: in $x$ direction, the cloud is
smaller and in $y$ direction it is larger than the fitted
Gaussian. This indicates that the expansion does not exactly follow a
scaling law of the form $n(r,y,z) = n_0(x/b_x,y/b_y,z/b_z)/b_xb_yb_z$
as assumed in previous analyses
\cite{Schaefer2010,Cao2011,Elliott2014a,Elliott2014b,Joseph2014}.
This is not very surprising since one would expect that at the
beginning of the expansion, the central part of the cloud follows
viscous hydrodynamics and becomes ballistic at later times, while the
dilute peripheral part of the cloud expands ballistically from the
beginning.

\begin{figure}
  \includegraphics[width=8cm]{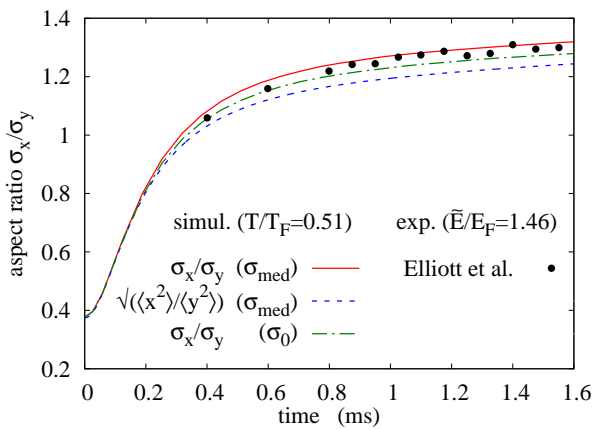}
  \caption{Time dependence of the aspect ratio $\sigma_x/\sigma_y$ of
    the expanding cloud in the simulation of the experiment by Elliott
    et al. \cite{Elliott2014a} with $\tilde{E}/E_F = 1.46$.
    \label{Fig:compareElliott}}
\end{figure}

The time dependence of the aspect ratio of the expanding cloud, i.e.,
the ratio of the fitted cloud widths $\sigma_x/\sigma_y$, is displayed
in \Fig{Fig:compareElliott}
as the solid line. Our result agrees very well with the experimental
data. For comparison, we also show as the dashed line what one obtains
if one defines the aspect ratio as $\sqrt{\ave{x^2}/\ave{y^2}}$. The
discrepancy between the two curves comes from the ballistic
low-density part discussed above.

We also studied the influence of the in-medium effects on the
expansion. Apparently the effects of mean field and in-medium cross
section practically cancel each other if the simulations with and
without in-medium effects are made with the same initial temperature
$T/T_F$. However, the mean field is necessary to get the correct
relation between the experimental observable $\tilde{E}/E_F$ and the
temperature $T/T_F$. The dash-dot line in \Fig{Fig:compareElliott}
represents the result for the aspect ratio obtained with mean field
but without in-medium effects in the cross section. Since the
collision rate with the free cross section $\sigma_0$ is lower than
the one with the in-medium cross section $\sigma_\med$, the inversion
of the aspect ratio is somewhat weaker than the one obtained in the
full calculation. However, the difference is too small to make a
conclusive statement whether the result with or without in-medium
cross section is in better agreement with the experimental data.

\begin{figure}
  \includegraphics[width=8cm]{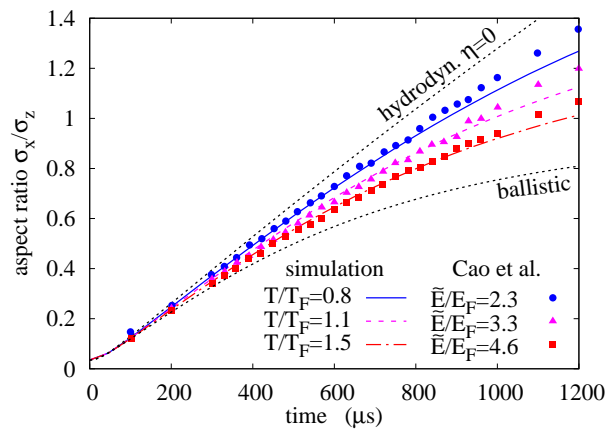}
  \caption{Time dependence of the aspect ratio $\sigma_x/\sigma_z$ of
    the expanding cloud in the simulation of the experiments by Cao et
    al. \cite{Cao2011} with $\tilde{E}/E_F = 2.3$, $3.3$, and $4.6$.
    \label{Fig:compareCao}}
\end{figure}

In \Fig{Fig:compareCao}
we show results for the aspect ratios in simulations (with mean field
and in-medium cross section) starting from a different initial
geometry and higher temperatures \cite{Cao2011}. For comparison, the
dotted lines show the limiting cases of an ideal hydrodynamic ($\eta =
0$) and a purely ballistic expansion. One observes that with
increasing temperature, the behavior of the system becomes less
hydrodynamic and approaches the collisionless regime. The same
behavior is seen in the experimental data. Although the agreement
between simulation and experiment is not as good as in the case
$\tilde{E}/E_F = 1.46$, it is satisfactory for a theory which does not
have any adjustable parameter.

Since in the case of the expansion the calculation is quite simple
(one simulation per expansion), we were in this case able to increase
the number of test particles and to reduce the time step by factors of
two (cf. \Table{tab:numericalparameters}). However, already with the
smaller number of test particles and the larger time step we obtained
practically the same results as shown in \Figs{Fig:compareElliott} and
\ref{Fig:compareCao}, only in the lower row of \Fig{Fig:expansion} the
higher number of test particles shows up in reduced statistical
fluctuations around the fitted density profile near the center of the
cloud.

\subsection{Quadrupole mode}
Let us now study the radial quadrupole oscillation. We consider a
trapped gas in the unitary limit with the same characteristics as in
the experiment of \Ref{Riedl2008}, see \Table{tab:expinnsbruck}.
The quadrupole mode is then excited at $t=0$ by giving all particles a kick
\begin{equation}
p_x \to p_x + c_Q x\,,\qquad p_y \to p_y - c_Q y\,,
\label{eq:kick}
\end{equation}
(we use $c_Q = 0.2 m\omega_r$) and we compute as observable the
quadrupole moment
\begin{equation}
Q(t) = \ave{x^2-y^2}(t)\,.
\end{equation}
To extract the frequency $\omQ$ and damping rate $\GamQ$ of the
quadrupole mode from the response $Q(t)$, we fit it with a function of
the form
\begin{equation} \label{eq:quad_fit}
  Q_{\fit}(t) = A e^{-\GamQ t} \sin \omQ t
  + B (e^{-\GamQ t} \cos \omQ t - e^{-\Gamma_1 t})\,.
\end{equation}
The motivation for the choice of this form, including the
non-oscillating exponential with decay constant $\Gamma_1$, is
two-fold: it is similar to the fit function used in the experimental
analysis \cite{Altmeyer2007}, and it coincides with the form of the
response one obtains analytically within the first-order moment method
\cite{Lepers2010}.

\begin{figure}
  \begin{center}
    \includegraphics{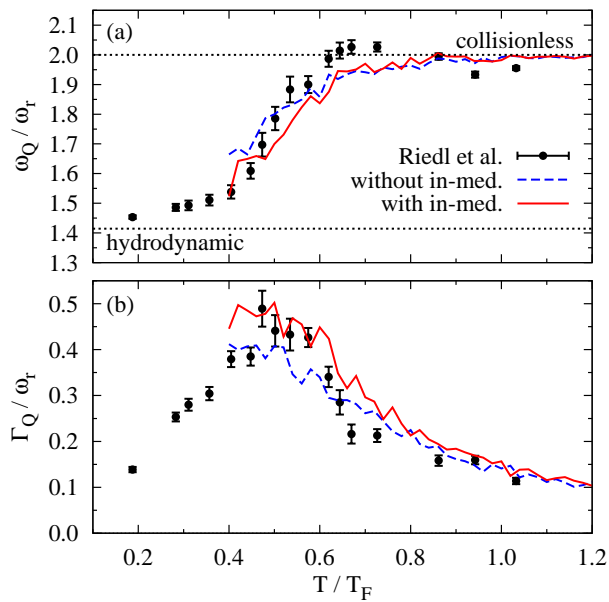}
  \end{center}
  \caption{Frequency (a) and damping (b) of the quadrupole mode in a
    system corresponding to the experiment by Riedl et
    al. \cite{Riedl2008}. In the simulations, we use $1/k_Fa = -0.1$
    instead of the unitary limit to avoid the divergence of the cross
    section for particles moving in parallel at the same speed. The
    dotted horizontal lines at $\omQ/\omr = \sqrt{2}$ and $2$
    correspond to the frequencies of the quadrupole mode in the
    hydrodynamic collisionless limits, respectively.
  \label{fig:quad}}
\end{figure}

\Figure{fig:quad}
shows the temperature dependence of the frequency (a) and the damping
rate (b) of the radial quadrupole mode of the gas whose parameters are
given in \Table{tab:expinnsbruck}. Results obtained with and without
in-medium effects ($\sigma_\med$, $U$) are shown as solid and dashed
lines, respectively. The data points are the experimental results from
\Ref{Riedl2008}. The irregularities of our results give an idea of the
numerical error. Because of the finite number of test particles, the
results for $Q(t)$ are somewhat noisy, especially at large times when
the amplitude of the mode has strongly decreased. This leads to some
uncertainties in the fitted values of $\omQ$ and $\GamQ$. Our
calculations are globally in good agreement with the data.

The influence of the in-medium effects is surprisingly weak. This is
in contrast with \cite{Riedl2008,Chiacchiera2009} where it was found
within the second-order moments method that with in-medium effects,
the collisionless regime was reached at too high temperature. Later it
was shown that compared with a numerical solution of the Boltzmann
equation, the second-order moments method tends to overestimate the
collision effects \cite{Lepers2010}, and that by extending the moments
method to fourth order the discrepancy between the results with
in-medium cross section and the experimental data was somewhat reduced
\cite{Chiacchiera2011}. If one assumes that the moments method
converges to the full solution of the Boltzmann equation, the present
results suggest that even higher orders in the moments method would
further reduce the influence of the in-medium modification of
the cross section.

However, we see that at $T/T_F \sim 0.4$, the in-medium effects lead
to a somewhat lower frequency, which is closer to the data and to the
hydrodynamic result $\omQ/\omr \to \sqrt{2}$. Furthermore, the maximum
damping near $T/T_F = 0.5$ is stronger and in better agreement with
the data if in-medium effects are included. Unfortunately, the mean
field $U$ cannot explain the experimental frequencies $\omQ/\omr > 2$
around $T/T_F=0.7$, although an attractive mean field is able to
increase the frequency of the quadrupole mode in the collisionless
limit to values above $2\omr$
\cite{Menotti2002,Massignan2005,Chiacchiera2009}.

Since our numerical solution of the Boltzmann equation has the
necessary flexibility, we can go a step further in our analysis and
simulate more closely the experimental procedure. To that end, we
excite the mode as before and let the system oscillate for some time
$t_\osc$, then we switch the trap off and let the system expand during
$t_\tof = 6/\ombar \approx 2$ ms \cite{Altmeyer2007}. Then we
calculate the quadrupole moment after the expansion,
$Q(t_\osc+t_\tof)$. We repeat the procedure for 50 different values of
$\ombar t_\osc$ from $0.04$ to $2.98$ (corresponding to $t_\osc
\approx 0\dots 1$ ms). Therefore, the whole calculation is very time
consuming.

\begin{figure}
  \begin{center}
    \includegraphics{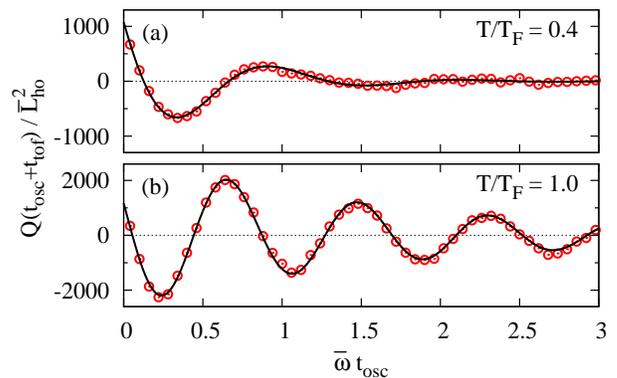}
  \end{center}
  \caption{Quadrupole moment after oscillation during $t_\osc$ and
    subsequent expansion during $t_\tof = 6/\ombar$, for two different
    initial temperatures $T/T_F = 0.4$ (a) and $1.0$ (b). The circles
    show the results of the simulation with in-medium effects, while
    the solid line corresponds to the fit by \Eq{eq:quad_tof_fit}. The
    system parameters are the same as in \Fig{fig:quad}.
  \label{fig:quad_exp_example}}
\end{figure}

Examples for the quadrupole moment after expansion as a function of
$t_\osc$ are shown in \Fig{fig:quad_exp_example}
(circles) for two different initial temperatures. In contrast to the
case without expansion \cite{Lepers2010}, we see that $Q$ does not
vanish for $t_\osc = 0$. The reason is easy to understand: The
excitation by the kick (\ref{eq:kick}) does not immediately create a
quadrupole moment, but a quadrupolar velocity field. During the
expansion, this is transformed into a quadrupole moment. Therefore, if
we want to determine the frequency $\omQ$ and damping $\GamQ$, we have
to modify the fit function and use instead of \Eq{eq:quad_fit} the
more general form \cite{Altmeyer2007}
\begin{equation} \label{eq:quad_tof_fit}
  Q_{\fit}(t) = A e^{-\GamQ t} \sin (\omQ t+\phi) + B e^{-\Gamma_1 t}\,.
\end{equation}
The fitted functions are also shown in
\Fig{fig:quad_exp_example}. One can see that, because of the
strong damping, the response for $T/T_F = 0.4$ (a) shows only a single
oscillation and is dominated by numerical noise for $t_\osc /\ombar
\gtrsim 1.5$. The determination of $\omQ$ and $\GamQ$ by the fit with
\Eq{eq:quad_tof_fit} is thus not very precise in this case (as
mentioned before, the same problem limits also the precision of $\omQ$
and $\GamQ$ in the case without expansion). At $T/T_F= 1$ (b), the
damping is weaker, and we get a higher precision.

\begin{figure}
  \begin{center}
    \includegraphics{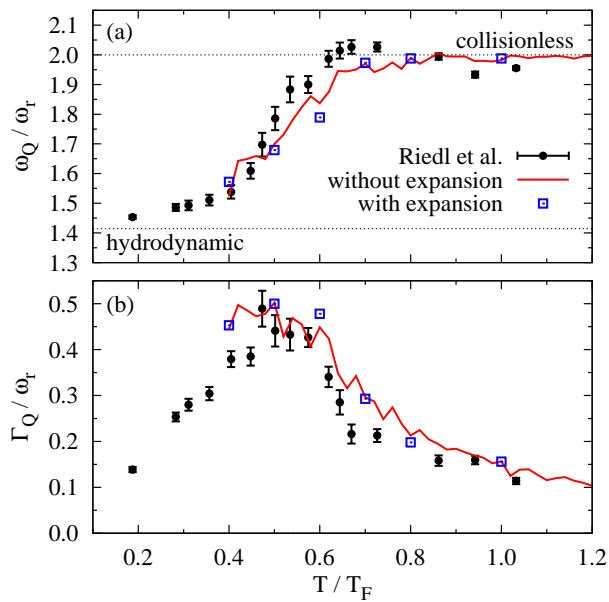}
  \end{center}
  \caption{Same as \Fig{fig:quad}, but the results were obtained from
    the quadrupole moment with (squares) and without (solid lines)
    expansion (both include in-medium effects).}
  \label{fig:quad_exp}
\end{figure}

We repeat this procedure for a couple of temperatures. The results for
$\omQ$ and $\GamQ$ are displayed as the squares in \Fig{fig:quad_exp}.
Comparing these results with those obtained without expansion (solid
lines), one concludes that the expansion before the measurement of the
quadrupole moment does not affect the extracted frequencies and
damping rates. There is one temperature where some discrepancy is
visible, namely $T/T_F = 0.6$, but as discussed before, the precise
extraction of the values $\omQ$ and $\GamQ$ is difficult in this
region of strong damping.

\section{Conclusions}\label{sec:conclusion}
In this paper, we described an extension of the test-particle method
of \Ref{Lepers2010} for the numerical solution of the Boltzmann
equation for ultracold trapped Fermi gases in the normal phase. The
main improvements are the possibility to treat realistic particle
numbers in realistic (strongly elongated) trap geometries and the
inclusion of the mean field and of the in-medium-modifications of the
cross section, calculated within a T matrix approximation. This method
allows us to simulate several recent experiments on the
non-equilibrium dynamics of unpolarized Fermi gases in the
normal-fluid phase.

As a first example, we studied the anisotropic expansion of a unitary
Fermi gas as measured in \Refs{Cao2011,Elliott2014a} for temperatures
$T/T_F \geq 0.4$. Without any adjustable parameter, we obtain an
excellent agreement with the time dependence of the aspect ratio
measured in \Ref{Elliott2014a} and an overall reasonable agreement with
the data of \Ref{Cao2011}. We notice that the peripheral region of the
cloud is never hydrodynamic and does not follow the scaling law which
approximately describes the expansion.

Then we discussed the radial quadrupole mode as measured in
\Ref{Riedl2008}. Again, we find that the numerical solution of the
Boltzmann equation gives a reasonably good description of the data. In
particular, the influence of in-medium effects on the frequency and
damping rate of the quadrupole mode seems to be much weaker than in
approximate solutions of the Boltzmann equation based on the method of
phase-space moments, where it was found to deteriorate the agreement
with the data \cite{Riedl2008,Chiacchiera2009}. We also show that the
expansion before the measurement of the deformation of the cloud does
not affect the final results for frequency and damping.

These examples show that, if solved correctly, the Boltzmann equation
is able to describe out-of equilibrium processes of unpolarized Fermi
gases, even in the strongly interacting regime (unitary limit), at a
quantitative level, at least at not too low temperatures. Note that
even at $T/T_F = 0.4$, only in a very small part of the trapped gas
the temperature is close to the local $T_c$ where one would expect
pseudogap effects to invalidate the Boltzmann equation. Of course, the
Boltzmann equation is not applicable below the superfluid transition
temperature ($\sim 0.3\,T_F$ for a trapped gas).

Since our Boltzmann code reproduces the anistroptic expansion
experiments of \Refs{Cao2011,Elliott2014a}, it would be interesting to
adapt it to calculate directly the temperature dependence of the shear
viscosity $\eta$, and to compare it with the one extracted from the
experiments under relatively strong assumptions (a refined analysis
was made in \Ref{Joseph2014}). In principle, the result should be
similar to that of \Refs{BruunSmith2005,BluhmSchaefer2014}, where the
viscosity was also calculated from the Boltzmann equation with
in-medium effects.

A challenging subject for future work would be the description of
shock waves as studied experimentally in \cite{Joseph2011} by
splitting the initial cloud into two and letting them collide in $z$
direction. The difficulty of this problem compared to the phenomena
studied in the present work is the rapid change of the distribution
function on a short length scale in the direction of the long axis of
the trap.

Also a quantitative description of the collision of two spin-polarized
clouds with opposite polarization as in the experiment of
\Ref{Sommer2011} is still missing. On a qualitative level, it was
already shown in \Ref{Goulko2011} that the Boltzmann equation
without in-medium effects can reproduce the different regimes
(bounce, transmission) that were experimentally observed. What
is particularly difficult in this case is the inclusion of in-medium
effects, since the T matrix approximation used in the present work
fails in the polarized case. A step into this direction was undertaken
in \Ref{Pantel2014}.

\appendix

\section{Approximations for the in-medium cross section}
\label{app:xsection}
Depending on the temperature, we use two different strategies to
approximate the in-medium cross section.
\subsection{Low temperature}
At low temperature, $\sigma_\med$ has typically a peak near $q = k_F$ (where
$k_F$ is the Fermi momentum), especially for small total momenta $k$
(cf. Fig.~2 in \Ref{Chiacchiera2009}). At very large $k$ or $q$, it
approaches the free cross section $\sigma_0$. One might therefore try to
fit \Eq{eq:inmedxsec} with a function of the form
\begin{equation}
  \sigma_\fit(k,q) = \sigma_0(q) \left[ 1 + A e^{-k^2/k_0^2} \frac{1}{(q -
  q_0)^2 + q_1^2} \right],
\end{equation}
where the parameters $A, k_0, q_0$, and $q_1$ are functions depending
on $T$, $\mubar$ and $a$ to be determined (as mentioned above, we
consider only the unpolarized case, i.e., $\mu_\up = \mu_\down =
\mubar$). However, the problem with a fit is that it is not clear how
the different $k$ and $q$ should be weighted; furthermore, the
obtained parameters $A, \dots, q_1$ as functions of $T$, $\mubar$ and
$a$ are too noisy to be interpolated.

We therefore follow a slightly different strategy. As a measure for
the quality of the parameterization, one can take its capability to
reproduce certain physical properties that depend on the
cross-section. For instance, one wishes to reproduce the equilibrium
values of the collision rate per particle, $\gamma$, or of the
viscous relaxation time, $\tau$.

Thus, instead of fitting directly $\sigma_\med$ with $\sigma_\fit$, we
determine the four parameters of $\sigma_\fit$ such as to reproduce
four different moments (i.e., integrals weighted with different powers
of $k$ and $q$) of $\sigma_\med$,
\begin{equation}
  m^\med_j = \int dk \int dq \sigma_\med(k,q) \kappa(k,q) \phi_j(k,q)\,,
\end{equation}
where $\kappa$ reads
\begin{equation}
  \kappa(k,q) = \left[ \frac{\atanh \left( \tanh \frac{X}{2} \tanh
  \frac{Y}{2} \right)}{\sinh X} \right]^2 ,
\end{equation}
with $X = 2 \beta \omega$ and $Y = \beta k q/2m$, and the four weight
functions $\phi_j$ ($j\in \{1,\dots, 4\}$) are
\begin{equation}
  \phi_1 = q\,,\quad \phi_2 = q^3\,,\quad 
  \phi_3 = q^5\,,\quad \phi_4 = k^2q^3\,.
\end{equation}
The four parameters $A\dots q_1$ are now determined from the four
equations $m^\med_j = m^\fit_j$ (the $m^\fit_j$ have of course the
same expressions as $m^\med_j$, only the cross section has to be
changed accordingly). Since $m^\med_1$ is exactly related to
$\gamma$ [cf. \Eq{eq:collrate}], and $m^\med_4$ is similar to the
integral appearing in the calculation of $\tau$
(cf.\ \Ref{Chiacchiera2009}), we are sure that $\sigma_\fit$ will give
good results for $\gamma$ and $\tau$.
\subsection{High temperature}
When the temperature increases, the shape of $\sigma_\med$ becomes
flatter and the Lorentzian part of $\sigma_\fit$ does not fit the real
behavior any more. At high temperature, the chemical potential becomes
negative and the Fermi functions in \Eq{eq:J} can be expanded as
$\fz(\xi) = e^{-\beta\xi}-e^{-2\beta\xi}+e^{-3\beta\xi}-\dots$. As a
consequence, the expansion of $J$ reads
\begin{multline}
  J(\omega,\kv;T) \approx J_\highT(\omega,\kv;T) - J_\highT(\omega,\kv;T/2) \\
  + J_\highT(\omega,\kv;T/3) - \dots\,.
\end{multline}
The result for high-temperature limit $J_\highT$ can be written in
closed form as
\begin{multline}
  J_\highT(\omega,\kv;T) = \frac{m^2 e^{\beta\mubar}}{2\pi\beta k}
   \Bigl[\frac{2}{\sqrt{\pi}}[F(X_+) - F(X_-)]\\ 
    + i \left(e^{-X_-^2} - e^{-X_+^2}\right)\Bigr],
\end{multline}
where $X_\pm = (q\pm k/2)/\sqrt{2mT}$ and $F$ denotes Dawson's
integral \cite{AbramowitzStegun} (see Appendix \ref{app:Dawson}).

In practice, we use the first three terms of the high-temperature
expansion in the case $\mubar < 0$, and the parameterization
$\sigma_\fit$ otherwise.
\section{Approximation for Dawson's integral}
\label{app:Dawson}
In the high-temperature expansion of the in-medium cross section there
appears a function $F(x)$, called Dawson's integral
\cite{AbramowitzStegun} and defined by
\begin{equation}
F(x) = e^{-x^2} \int_0^x e^{t^2} dt\,.
\end{equation}
Since speed is more important than precision for our purpose, we
approximate $F(x)$ for $-\infty < x < \infty$ by a simple rational
function of the form
\begin{equation}
F(x) \approx \frac{x+a_1 x^3+a_2 x^5}{1+b_1 x^2+b_2 x^4+2 a_2 x^6}\,,
\end{equation} 
which automatically reproduces the asymptotic behavior of $F(x)$ for
$x\to 0$ and $x\to\pm\infty$. By minimizing the maximum relative
error, we obtain the following values for the parameters:
\begin{align}
a_1 &= 0.133931\,,  & b_1 &= 0.853463\,,\nonumber\\
a_2 &= 0.0989404\,, & b_2 &= 0.228679\,.
\end{align}
The maximum relative error of this approximation is $0.53\%$.

\end{document}